\documentclass{article}

\usepackage{microtype}
\usepackage{graphicx}
\usepackage{subfigure}
\usepackage{booktabs} 
\usepackage{amsmath}
\usepackage{xspace}

\usepackage{bbding}
\usepackage{pifont}
\usepackage{wasysym}
\usepackage{amssymb}

\newcommand{\ourwork}[0]{MEADOW\xspace}

\newcommand{\tsup}[1]{\textsuperscript{\texttt{#1}}}
\newcommand{\tsub}[1]{\textsubscript{\texttt{#1}}}

\newcommand{\ttsub}[2]{\large\texttt{#1}\tsub{#2}\normalsize\xspace}
\newcommand{\ttsup}[2]{\large\texttt{#1}\tsup{#2}\normalsize\xspace}
\newcommand{\ttt}[1]{\large\texttt{#1}\normalsize\xspace}

\newcommand{\smqktv}[0]{\ttt{Q+SM(QK\tsup{T})xV}}
\usepackage{hyperref}

\newcommand{\isaccepted}{}

\usepackage{mlsys2025}

\mlsystitlerunning{MEADOW: Memory-efficient Dataflow and Data Packing for Low Power Edge LLMs}

\begin{document}

\twocolumn[
\mlsystitle{MEADOW: Memory-efficient Dataflow and Data Packing for Low Power Edge LLMs}



\mlsyssetsymbol{equal}{*}
\begin{mlsysauthorlist}
\mlsysauthor{Abhishek Moitra}{yale}
\mlsysauthor{Arkapravo Ghosh}{yale}
\mlsysauthor{Shrey Agarwal}{iitr}
\mlsysauthor{Aporva Amarnath}{ibm}
\mlsysauthor{Karthik Swaminathan}{ibm}
\mlsysauthor{Priyadarshini Panda}{yale}
\end{mlsysauthorlist}

\mlsysaffiliation{yale}{Department of Electrical and Computer Engineering, Yale University, CT, USA}
\mlsysaffiliation{iitr}{IIT Roorkie, Roorkie, India}
\mlsysaffiliation{ibm}{IBM Research - Yorktown Heights Yorktown Heights, NY USA}

\mlsyscorrespondingauthor{Abhishek Moitra}{abhishek.moitra@yale.edu}

\mlsyskeywords{Machine Learning, MLSys}

\vskip 0.3in

\begin{abstract}

The computational and memory challenges of large language models (LLMs) have sparked several optimization approaches towards their efficient implementation. 
While prior LLM-targeted quantization, and prior works on sparse acceleration have significantly mitigated the memory and computation bottleneck, they do so assuming high power platforms such as GPUs and server-class FPGAs with large off-chip memory bandwidths and employ a generalized matrix multiplication (GEMM) execution of all the layers in the decoder. In such a GEMM-based execution, data is fetched from an off-chip memory, computed and stored back. However, at reduced off-chip memory capacities, as is the case with low-power edge devices, this implementation strategy significantly increases the attention computation latency owing to the repeated storage and fetch of large intermediate tokens to and from the off-chip memory. Moreover, fetching the weight matrices from a bandwidth constrained memory further aggravates the memory bottleneck problem. To this end, we introduce \ourwork, a framework that significantly reduces the off-chip memory access for LLMs with a novel token-parallel head-sequential (TPHS) dataflow. Additionally, \ourwork applies weight packing, that performs loss-less decomposition of large weight matrices to their unique elements thereby, reducing the enormous weight fetch latency. \ourwork demonstrates 1.5$\times$ and 2.5 $\times$ lower decode and prefill latency, respectively, compared to a GEMM-based LLM implementation on the low power Xilinx ZCU102 FPGA platform that consumes less than 10W. Additionally, \ourwork achieves an end-to-end latency improvement of over 40\%, compared to prior LLM optimization works.

\end{abstract}
]



\printAffiliationsAndNotice{}  


\section{Introduction}
\label{sec:introduction}


The explosive growth of large language models (LLMs) necessitates efficient, low-power hardware solutions to make them accessible across diverse AI applications \cite{zhang2024llmcompass, minaee2024large, chang2024survey}. 
In particular, there have been several efforts to deploy LLMs across a swath of applications at the edge, ranging from autonomous driving systems~\cite{marcu2023lingoqa} to mobile device assistants~\cite{mobileAIBench}. Even though there have been a few custom ASIC solutions targeting fixed transformer models \cite{tambe202322, park2024lpddr}, their significant design/verification complexity and the consequent impact on the time-to-market makes it difficult for them to cater to the rapidly changing nature of the models and their underlying applications. On the other hand, more general-purpose CPU/GPU/TPU solutions deployed on the cloud cannot be replicated on edge devices due to their inherent Size, Weight and Power (SWaP) limitations.

Data-center scale hardware solutions, like the AMD Alveo series ~\cite{alveo}, leverage high bandwidth memory (HBM) to handle the intense demands of LLMs, but they also consume over 200 Watts of power. In contrast, platforms like the Xilinx ZCU102 \cite{zcu102} and Xilinx ZCU104 \cite{zcu104} offer a reconfigurable, low-power alternative with a sub-10 Watt power budget, making them well-suited for exploring the extensive design space of LLMs, while, balancing power and performance. However, without HBM, these platforms face limitations in available memory bandwidth. This constraint presents a challenge, as the attention computations that drive modern LLMs are highly memory-bound. To mitigate the memory bottleneck in LLMs, techniques like weight quantization \cite{xiao2023smoothquant, lin2024awq, xu2024llamaf} and sparse computation \cite{huang2024elsa, zhang2024llmcompass} have been proposed to reduce data transfer and computational complexity \cite{wang2023cta, ma2023llm}. However, these solutions are largely tailored for larger GPUs and/or TPUs. Achieving efficient LLM acceleration on low power-budget devices with restricted memory, calls for a cohesive approach that combines architecture optimization, dataflow restructuring, and parameter compression.

\begin{figure*}[h!]
    \centering
    \includegraphics[width=\textwidth]{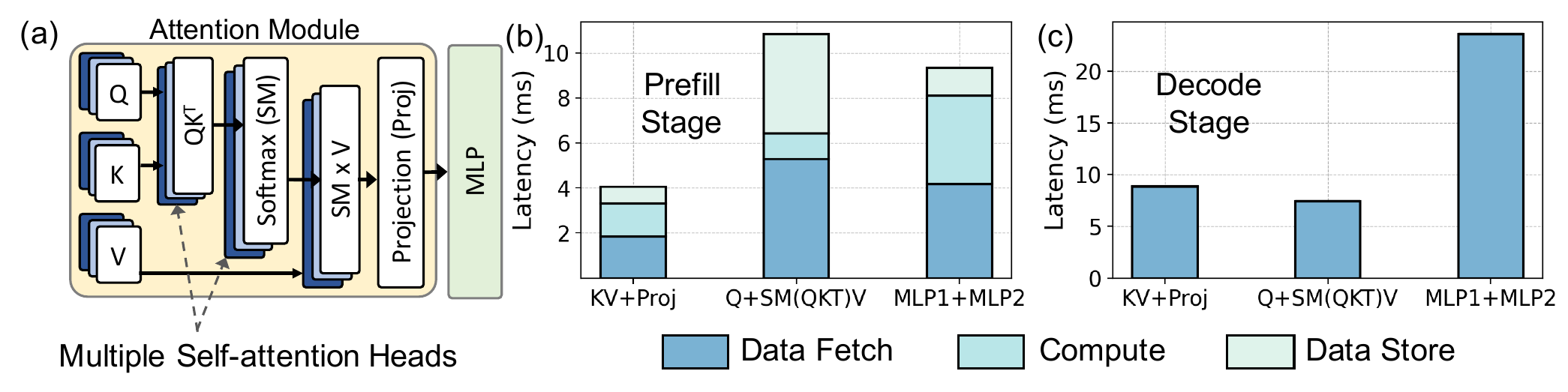}\vspace{-5mm}
    \caption{Figure showing the (a) Decoder architecture used in LLMs (b) the prefill latency distribution across data fetch, store and computation across different layers in the decoder (c) the decode latency distributions. During decode, compute and storage latency is negligible compared to the weight and input fetch latency. All latency results are based on OPT-125M LLM implementation on the Xilinx ZCU102 FPGA with off-chip DRAM bandwidth = 12Gbps.}
    \label{fig:intro_fig}
\end{figure*}


A typical LLM, especially for generative language processing, comprises of multiple layers of decoder hierarchy. A decoder architecture (as shown in Fig.~\ref{fig:intro_fig}a) has self-attention and matrix multiplication operations. During inference, the LLM operates in two stages: \emph{Prefill} and \emph{Decode}. In the prefill stage, a user-provided prompt is decomposed into multiple tokens. These tokens simultaneously undergo matrix multiplications with multi-dimensional weights to yield \ttt{Q}, \ttt{K} and \ttt{V} outputs. Subsequently, the \ttt{Q}, \ttt{K} and \ttt{V} values undergo fine-grained spatial correlations by means of \ttt{SM}(\ttsup{QK}{T})\ttt{xV} operations in multiple self-attention heads, where \ttt{SM} denotes a Softmax operation. The attention outputs are finally projected to higher dimension space by the projection (\texttt{Proj}) and \ttt{MLP} layers. Post prefill, the LLM enters the decode stage, where it predicts subsequent output tokens one-by-one. 

In most prior works, the \ttt{SM}(\ttsup{QK}{T}) \ttt{xV} layers are executed in the form of a generalized matrix multiplication (GEMM) operation \cite{zeng2024flightllm, wang2023cta, huang2024elsa}. Here, the input matrices for each self-attention head are fetched from the off-chip memory, processed in the GEMM array, and the output is stored back to the off-chip memory. As shown in Fig.~\ref{fig:intro_fig}b, under limited off-chip memory bandwidth (12~Gbps), this repeated data transfer significantly increases the latency in the prefill stage, where larger input sizes exacerbate memory access demands. During the decode stage, however, the input size is much smaller, reducing compute and data storage overheads to a negligible fraction, while the weight fetches dominate latency (Fig. \ref{fig:intro_fig}c). Thus, optimizing on memory accesses through efficient compute dataflow in both the prefill and decode stages is essential to reduce the overall latency.

To address the above challenges, we introduce the \ourwork framework. During the prefill and decode stage, \ourwork executes the \ttt{KV}, \ttt{Proj} and \ttt{MLP} layers in the GEMM mode while, the \ttt{Q}, \ttsup{QK}{T}, \ttt{SM}, and \ttt{SMxV} layers are executed with a novel Token-Parallel Head-Sequential (TPHS) dataflow which performs effective layer pipelining and significantly reduces the off-chip data fetches and storage latency. To further mitigate the latency and bandwidth overhead of weight fetches, \ourwork implements Weight Packing, which compacts the weight matrix by transferring only its unique elements, significantly minimizing weight transfer volume. Additionally, \ourwork applies bit-packing techniques on the weights to maximize memory bandwidth utilization, enhancing memory efficiency. 


The key contributions of our work are as follows:
\begin{enumerate}
    \item We propose \ourwork that uses a novel Token Parallel Head Sequential (TPHS) dataflow to compute the \ttt{SM}(\ttsup{QK}{T})\ttt{xV} layers in pipeline, significantly reducing the volume of data transfers to and from off-chip memory.
    \item We introduce Weight Packing, a technique that decomposes LLM weight matrices into unique elements to minimize weight fetch latency at prefill and decode stages. Additionally, to further accelerate weight fetches and maximize DRAM bandwidth efficiency, we implement bit-packing to compactly store and transfer weight data. Weight packing is an approximation-less technique that yields loss-less accuracy performance.
    \item We evaluate \ourwork on the ZCU102 FPGA with a peak power budget of 10W across varying off-chip DRAM bandwidths and input token lengths on state-of-the-art OPT-125M and OPT-1.1B LLM models. \ourwork achieves 2.5$\times$ and 1.5$\times$ lower prefill and decode latency compared to GEMM-based implementations for 1-6 Gb/s data bandwidth ranges. \ourwork also achieves over 40\% end-to-end latency improvement compared to prior LLM optimization works. 
    \item We demonstrate the generalizability of \ourwork across vision transformer (ViT) benchmarks, achieving 1.6$\times$ lower inference latency compared to GEMM-based ViT implementations. We also demonstrate how \ourwork can be applied to multiple FPGA configurations with varying PE sizes and memory bandwidth. 
\end{enumerate}

\section{Related Work}

\textbf{Data compression techniques:} Weight and input quantization is a widely adopted approach for data compression in LLMs. Works such as SmoothQuant \cite{xiao2023smoothquant}, AWQ \cite{lin2024awq}, and LlamaF \cite{xu2024llamaf} apply fake quantization methods to lower off-chip data transfers, and dequantize the compressed inputs and weights during computation to maintain good accuracy. A recent work MECLA \cite{qin2024mecla} applies a sub-matrix partitioning technique wherein, different sub-matrices within a larger matrix is approximated as a function of a base sub-matrix. 

\textbf{Sparse Computations:} 
Sparse computation techniques leverage the inherent dynamic sparsity of LLMs to reduce computation. Unstructured sparsity, as implemented in methods like ELSA \cite{huang2024elsa, fang2022algorithm, chen2023dynamic} with N:M sparsity, selectively prunes non-essential connections, effectively reducing computational load. FlightLLM \cite{zeng2024flightllm} implements N:M sparse computation using FPGA-based accelerators with HBM to address memory bottlenecks.

Structured pruning addresses the limitations of unstructured pruning by removing entire blocks or groups of computations. For example, token compression in CTA \cite{wang2023cta} reduces memory and compute demands by compressing less critical tokens. Gradient-based pruning, as used in LLM Pruner \cite{ma2023llm}, selectively prunes attention heads based on gradient information, focusing computational resources on essential parts of the model. ALISA \cite{zhao2024alisa} focuses on retaining tokens that are crucial towards generating new tokens via a sparse window attention technique. FACT \cite{qin2023fact} focuses on performing eager computation of attention tokens at minimal computation overhead and performing sparse computations for subsequent layers. 

\textbf{Parallel research directions} towards designing more hardware efficient transformer architectures are also being developed. EdgeBERT \cite{tambe2021edgebert}, PIVOT \cite{moitra2024pivot} and TReX \cite{moitra2024trex} use entropy of inputs to perform dynamic voltage scaling, attention skipping and reuse, respectively to achieve hardware efficiency. FlexLLM \cite{miao2024flexllm} introduce a unique inference and parameter-efficient finetuning to achieve efficient yet, highly accurate LLMs.

\ourwork is an orthogonal solution to prior techniques, introducing architectural and dataflow innovations along with weight packing to optimize weight fetch latency. By restructuring the dataflow and enhancing memory access patterns, \ourwork minimizes latency in retrieving weights, addressing memory bottlenecks in low memory bandwidth hardware without sacrificing model accuracy.

\section{\ourwork Architecture}
\label{sec:meadow_arch}

\begin{figure*}[h!]
    \centering
    \includegraphics[width=\textwidth]{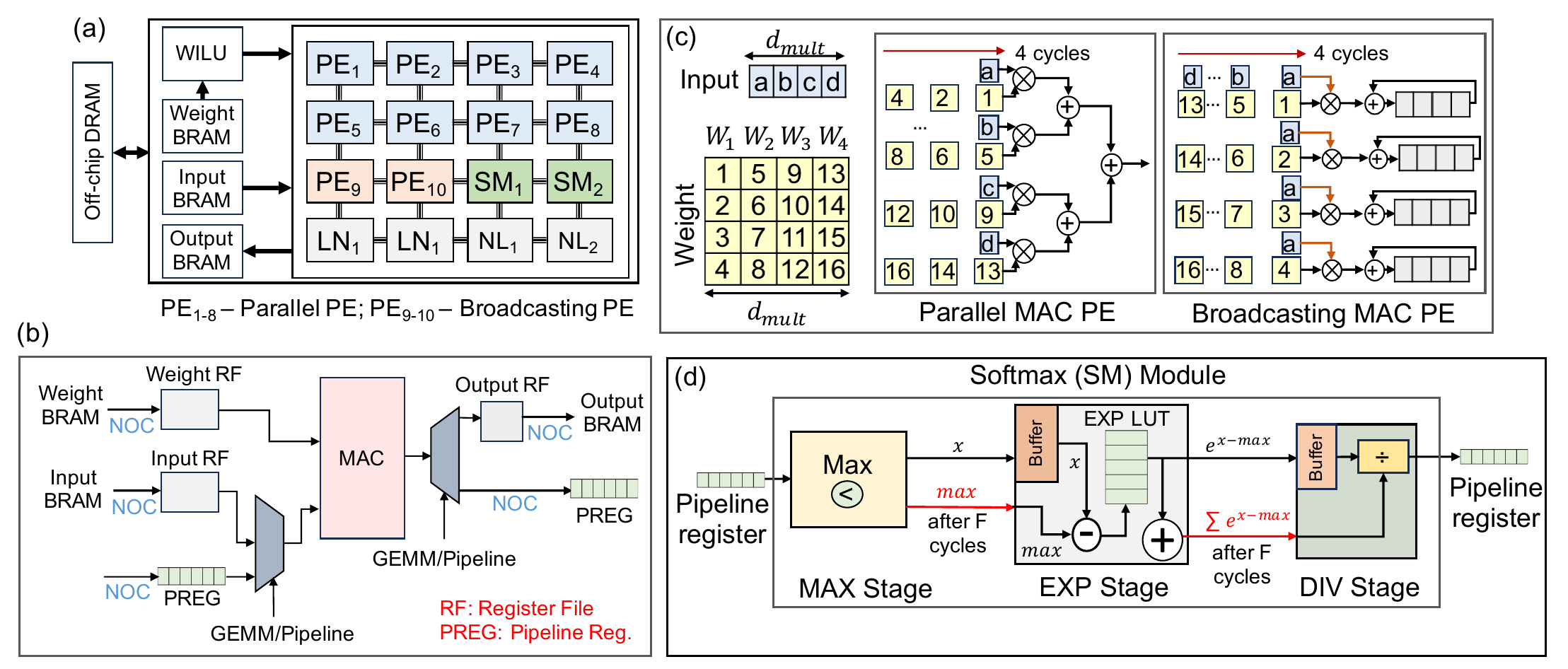}\vspace{-5mm}
    \caption{(a) Tiled architecture of \ourwork containing parallel and broadcasting processing elements (PEs), pipelined softmax (SM) module, modules for layer normalization (LN) and non-linear activation functions like ReLU/GeLU (NL). (b) The hybrid PE architecture capable of operating in GEMM and pipelined modes. (c) Architecture and execution flow of a parallel and broadcasting MAC PE. (d) The pipelined softmax (SM) module. }
    \label{fig:arch_overview}
\end{figure*}

\ourwork follows a tiled architecture as shown in Fig. \ref{fig:arch_overview}a containing multiple processing elements (PEs), modules for layer normalization (LN), softmax operations (SM) and non-linear (NL) activation functions, such as, ReLU/GeLU. Each PE contains several multipliers and accumulators to carry out the multiply-accumulate operations. For computation, the input data is fetched from the off-chip DRAM to the Input block RAM (BRAM). The raw input values are directly transferred to the respective input register files (RF) of the PEs. Since, \ourwork applies an additional weight packing to reduce off-chip weight fetches, the Weight BRAM stores the packed and encoded weight values which first needs to be processed by the Weight Unpacking and Index Look-up (WILU) Module. The WILU module reads data from the Weight BRAM and sends the data to the respective weight RFs of the PEs. The outputs from each PE are stored back to the output BRAM. All communications between BRAM and PE, SM, LN and NL modules are enabled by the network on chip (NoC) interconnect. The NoC additionally handles data communication between PEs and SM modules to facilitate the TPHS dataflow, defined in Section~\ref{sec:tphs}.



\textbf{Hybrid PE for GEMM and Pipelined Execution:} \ourwork employs a dual execution strategy: GEMM mode for the \ttt{KV}, \ttt{Proj}, and \ttt{MLP} layers, and the TPHS dataflow for the \ttt{Q}, \ttsup{QK}{T}, \ttt{SM}, and \ttt{SMxV} layers, enabling pipelined execution that minimizes data fetch and store latency. To support both GEMM and pipelined modes seamlessly, \ourwork utilizes a hybrid PE architecture, designed for flexible execution across modes. The PE shown in Fig. \ref{fig:arch_overview}b, integrates a multiply-accumulate (MAC) unit, input, weight, and output register files (RF), along with a pipeline register (PREG). All RFs and the pipeline registers are double-buffered to minimize data fetch and store latency \cite{moitra2024pivot}.

For the GEMM mode, data from the input and weight BRAMs are loaded into the input and weight RF, respectively. These data values are fetched and processed in the MAC array and the outputs are stored in the output RF. Once the output RF reaches capacity, the data is transferred to the output BRAM via the NoC. In contrast, for the pipelined mode, weights are loaded from the BRAM into the weight RF while the inputs are fetched directly from the pipeline register. The Input BRAM remains inactive during the pipelined mode of operation. After the MAC operation, the outputs are transferred directly through the NoC to the pipeline register of a target module (such as the softmax unit or another PE) in the subsequent pipeline stage.


\textbf{Parallel and Broadcasting PE:} \ourwork's tiled architecture contains a mix of Parallel MAC and Broadcasting MAC PEs (for example \ttsub{PE}{1-8} = Parallel and \ttsub{PE}{9-10} are Broadcasting MAC PEs as shown in Fig. \ref{fig:arch_overview}a). As shown in Fig. \ref{fig:arch_overview}c, both parallel and broadcasting MAC PEs use an array of multipliers but use different accumulation strategies. The Parallel MAC PE incorporates an adder tree, allowing it to multiply all elements along the multiplication dimension (\ttsub{d}{mult}) in a single cycle. In contrast, the Broadcasting MAC PE features accumulators (registers coupled with adders), enabling it to broadcast each input element along \ttsub{d}{mult} across all corresponding output channels and perform multiplication and accumulation sequentially over \ttsub{d}{mult} cycles. The Parallel MAC and Broadcasting MAC PEs are essential for facilitating the TPHS dataflow, described in Section~\ref{sec:tphs}.

\textbf{Pipelined Softmax Module (SM Module):} The numerically stable softmax computation of a given token is shown in Equation \ref{eq:softmax}. 
\begin{equation}
    SM = \frac{e^{x_i-max}}{\Sigma_i e^{x_i-max}}
    \label{eq:softmax}
\end{equation}
The computation requires three sequential stages: 1) finding the maximum across all the features in the token, 2) computing the exponent and the summation of all exponents and, 3) f
inally, dividing each exponent value with the exponent summation. Due to the sequential nature of the softmax stages, it is latency intensive. To this end, \ourwork pipelines the three stages across tokens to improve the softmax computation throughput. As shown in Fig. \ref{fig:arch_overview}d, the SM Module consists of three pipelined stages \ttt{MAX}, \ttt{EXP} and \ttt{DIV}. Each stage processes a token feature-by-feature over \ttt{F} cycles, where \ttt{F} is the number of features in the token. The \ttt{MAX} stage compares the feature values and returns the maximum value at the end of \ttt{F} cycles. Subsequently, the values are written to the \ttt{EXP} stage buffer. In the \ttt{EXP} stage, the maximum value output from the \ttt{MAX} stage is subtracted from each feature and the exponent values are computed. For hardware efficiency, the exponent is computed using the \ttt{EXP} \ttt{LUT} lookup table. Simultaneously, the exponent values are summed up and are stored in the DIV stage buffer. Finally, in the \ttt{DIV} stage the exponent values are fetched from the \ttt{DIV} stage buffer and divided by the exponent summation value.


\begin{figure*}[h!]
    \centering
    \includegraphics[width=\textwidth]{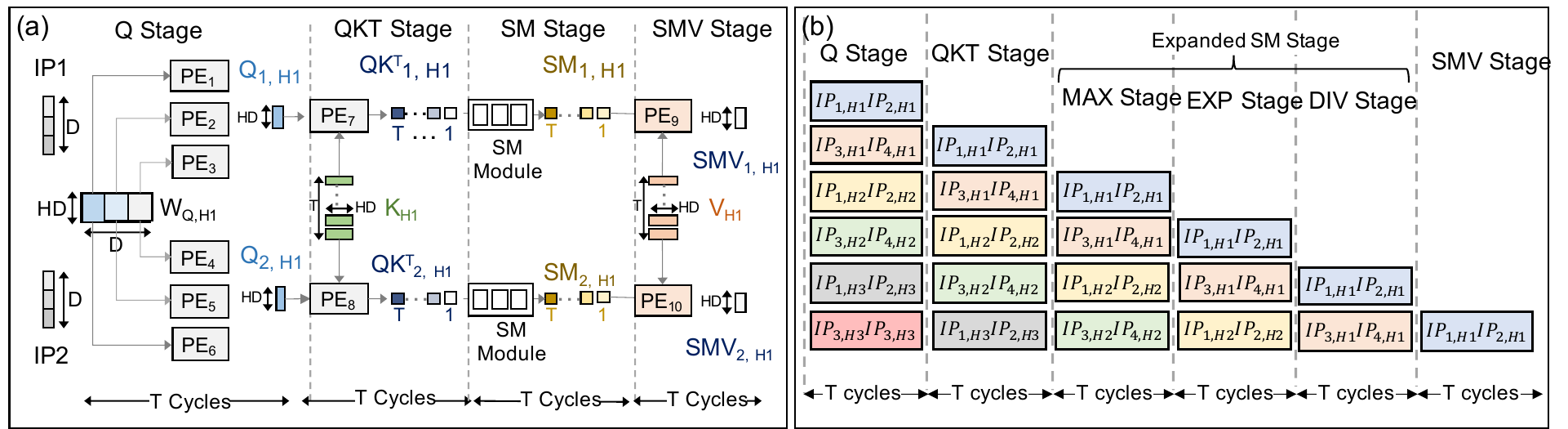}\vspace{-5mm}
    \caption{Figure showing an example of (a) token parallel head sequential (TPHS) dataflow with two input tokens being processed parallely (b) The pipelined execution of a transformer with 3 heads (\texttt{H1-H3}) and 4 input tokens (\ttsub{IP}{1-4}).}
    \label{fig:pipeline}
\end{figure*}

\section{TPHS Dataflow}\label{sec:tphs}

To overcome the memory bound implementation of \ttt{Q}, \ttsup{QK}{T}, \ttt{SM(QKT)} and \ttt{SMxV} operation, \ourwork uses the token parallel head sequential (TPHS) dataflow. The TPHS dataflow shown in Fig. \ref{fig:pipeline}a, pipelines all the computations for each attention head in parallel across multiple tokens. In the example in Fig. \ref{fig:pipeline}a, we show how attention head 1 (\ttt{H1}) is computed for input tokens \ttsub{IP}{1} and \ttsub{IP}{2}. The TPHS dataflow requires the following data from the off-chip DRAM to be stored before computation- the input tokens \ttsub{IP}{1} and \ttsub{IP}{2} of size \ttt{1xD} each, the \ttsub{K}{H1}, \ttsub{V}{H1} pre-computed values for head H1 of size \ttt{TxHD}, where \ttt{T} and \ttt{HD} are the total number of input tokens and head dimension, respectively. Additionally, for the \ttsub{Q}{H1} computation, the \ttsub{W}{Q,H1} matrix of dimensions \ttt{DxHD} are required.

\begin{figure*}[h!]
    \centering
    \includegraphics[width=0.95\textwidth]{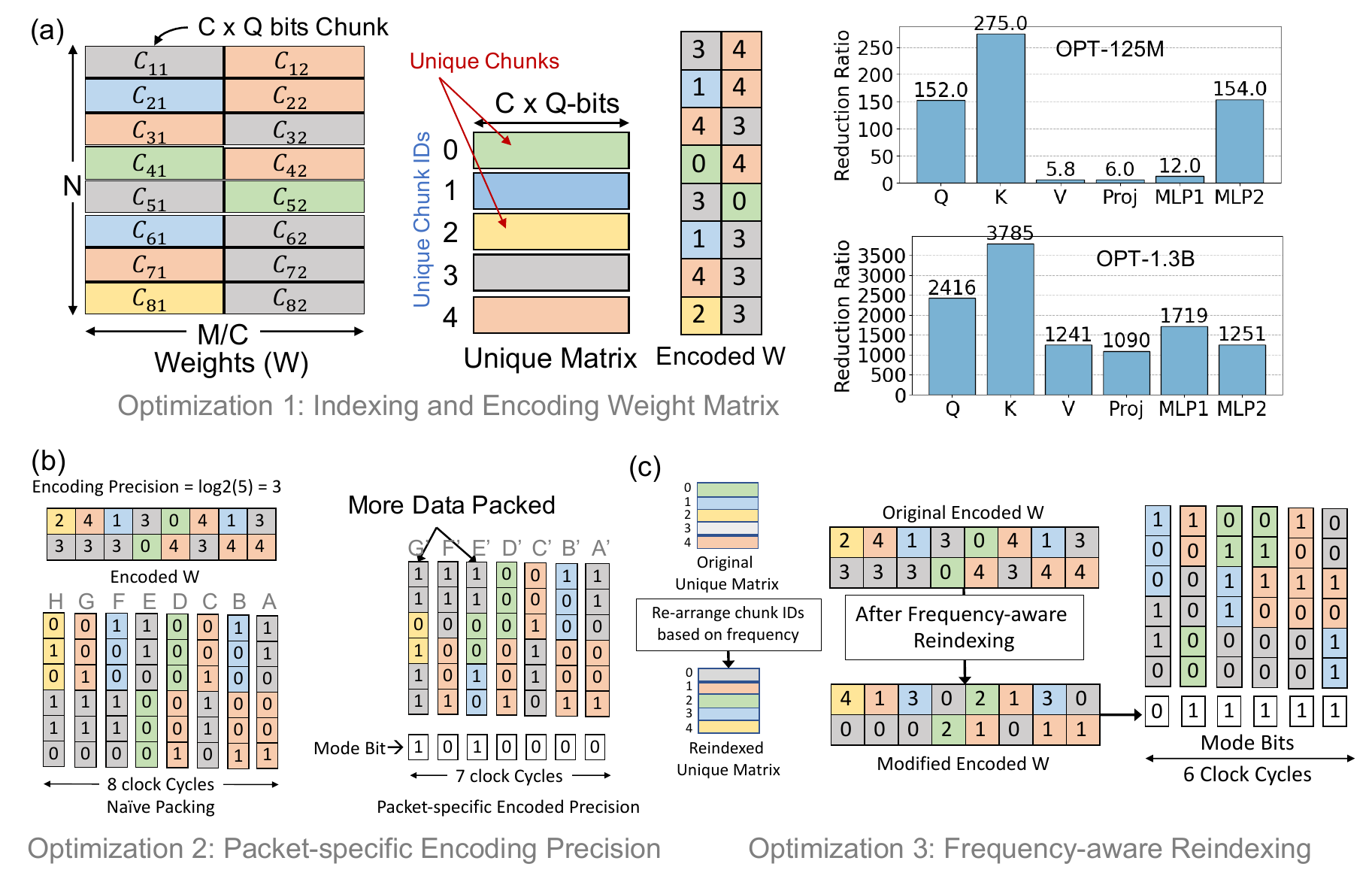}\vspace{-5mm}
    \caption{Figure showing (a) process of generating the unique matrix and the trends in the reduction ratios for OPT-125M and OPT-1.3B LLM models across different layers in the decoder. Reduction ratios are averaged across all the decoder layers. (b) packet-specific encoding precision and (c) frequency-aware reindexing to further optimize the DRAM bandwidth.}
    \label{fig:indexing_data_packing}
\end{figure*}
\ttsub{IP}{1} and \ttsub{IP}{2} are multiplied by \ttsub{W}{Q,H1} parallelly in \ttsub{PE}{1-3} and \ttsub{PE}{4-6}, respectively. This results in \ttsub{Q}{1,H1} and \ttsub{Q}{2,H1} for the two input tokens. \ttsub{Q}{1,H1} and \ttsub{Q}{2,H1} data is sent to the pipeline registers of \ttsub{PE}{7} and \ttsub{PE}{8}, respectively where they are multiplied with \ttt{T} tokens of \ttsub{K}{H1} resulting in \ttsub{\ttsup{QK}{T}}{1,H1} and \ttsub{\ttsup{QK}{T}}{2,H1} values over \ttt{T} cycles. 
At each cycle, the \ttsup{QK}{T} outputs are sent to the \ttt{MAX} stage of softmax module which returns the maximum across all the \ttsup{QK}{T} values at the end of \ttt{T} cycles. Subsequently, these values are forwarded to the \ttt{EXP} stage and the \ttt{DIV} stage which finally yield the \ttt{SM} values over \ttt{T} cycles. In Fig. \ref{fig:pipeline}a, the \ttt{MAX}, \ttt{EXP} and \ttt{DIV} stages are combined into \ttt{SM} stage for simple visualization. The respective softmax outputs are sent to the pipeline registers of the broadcasting PEs \ttsub{PE}{9} and \ttsub{PE}{10} to compute the \ttt{SMxV} output. Here, the \ttt{SM} outputs are multiplied with \ttsub{V}{H1} tokens over \ttt{T} cycles to yield \ttsub{SMV}{1,H1} and \ttsub{SMV}{2,H1} outputs for both tokens. The \ttsub{SMV}{1,H1} and \ttsub{SMV}{2,H1} outputs are stored to the off-chip DRAM. As shown in Fig. \ref{fig:pipeline}a, each stage requires \ttt{T} clock cycles. 

Fig. \ref{fig:pipeline}b shows an example of the pipelined execution of TPHS dataflow. Here, we consider a transformer having 3 self-attention heads with 4 tokens and two tokens being simulatenously processed. Additionally, we show the expanded stages inside the SM module for better visualization. In the TPHS dataflow, first all H1 self-attention heads are computed for every input token before proceeding to the computation of H2. This minimizes the amount of back-and-forth data transfers of the \ttsub{W}{Q}, \ttt{K} and \ttt{V} matrices thereby minimizing additional latency overhead. 

\section{Weight Packing}


\subsection{Creating the Unique Matrix}

Let \ttt{W} be a matrix of trained weight values with dimensions \ttt{NxM}, where \ttt{M} represents the inner product dimension. As shown in Fig. \ref{fig:indexing_data_packing}a, the inner dimension \ttt{M} is divided into chunks of size \ttt{C}, where each element in \ttt{C} is a Q-bit value based on the quantization of the weight matrix.
Next, as illustrated in Fig. \ref{fig:indexing_data_packing}a, a \texttt{Unique Matrix} is generated, containing the unique chunks, each assigned a unique ID. These chunk IDs are used to encode the weight matrix, resulting in the creation of the \texttt{Encoded W} matrix. To intuitively understand the amount of redundancy in the LLM weight matrices, we define the reduction ratio as the ratio between the total number of chunks in the encoded W matrix ($N\times M/C$) and the number of unique chunks. Higher reduction ratio signifies more redundancy and vice-versa. 
As seen in Fig. \ref{fig:indexing_data_packing}a, for the decoder weights of OPT-125M and OPT-1.3B the reduction ratio varies in the order of $10^2$ to $10^3$ suggesting high redundancy in the weight matrices.

\subsection{Packet-specific Encoding Precision}
To improve the DRAM bandwidth efficiency, multiple elements of the encoded W matrix are grouped together to form a packet and transferred from the DRAM for processing. 
As shown in Fig. \ref{fig:indexing_data_packing}b with naive data packing, all the packets use the same data precision to represent the encoded weights. The precision here is determined by the maximum number of unique chunks in the unique matrix (5 as in the Fig. \ref{fig:indexing_data_packing}a). However, using homogeneous bit-precision across packets lead to inefficiencies, as cycles are wasted transmitting low-precision encoded values that could otherwise be represented with fewer bits. For example, packets \ttt{E} and \ttt{F} use 3-bit precision to represent 2-bit numbers. 

To this end, we employ packet-specific bit-precision to represent the encoded values, where each packet is assigned an optimal precision to maximize packing efficiency. As depicted in Fig. \ref{fig:indexing_data_packing}b, employing packet-specific bit-precision allows low-bit encoded values to be packed together more effectively, thereby reducing the number of cycles required for transmission. 
The encoding precision for each packet is determined by the maximum encoded value in the respective packet. Additional mode bits are now used to determine the bit-precision of each packet (for example 3-bits for packet \ttt{A'} and 2-bits for packets \ttt{E'}, \ttt{G'}). Packets with mode = 0 and mode = 1 use 3-bits and 2-bits to represent the encoded values, respectively. The mode bits will be used by the WILU module to unpack the grouped encoded values. Packet-specific encoding allows packing more data per packet thereby improving the DRAM bandwidth efficiency.

\subsection{Frequency-aware Re-indexing}

As illustrated in Fig. \ref{fig:indexing_data_packing}c, frequently occurring chunk IDs in the encoded W matrix (e.g., chunk ID = 3) may necessitate higher precision, which can limit the efficiency of bit packing.
In frequency-aware re-indexing, the chunk IDs are re-assigned to each unique chunk based on their frequency of occurrence \textit{i.e.,} chunk IDs appearing more frequently are assigned lower chunk ID. For instance, in the example presented in Fig. \ref{fig:indexing_data_packing}c, chunk IDs [0, 1, 2, 3, 4] with frequencies [2, 2, 1, 6, 5] are re-assigned new chunk IDs [2, 3, 4, 0, 1]. This approach increases the proportion of low-precision chunk IDs in the encoded W matrix, resulting in efficient bit packing and thereby reducing transfer cycles. The modified encoded W and the reindexed unique matrix are transferred from the DRAM for processing.

\subsection{Weight unpacking and Index Look-up Module}
\label{sec:unpacking_hw}
\begin{figure}[h!]
    \centering
    \includegraphics[width=\linewidth]{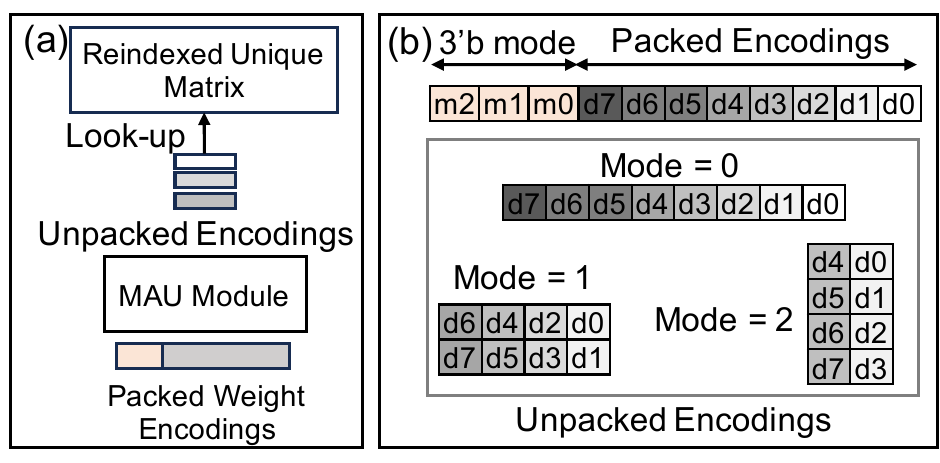}\vspace{-5mm}
    \caption{(a) The WILU Module (b) The mode-aware unpacking (MAU) module.}
    \label{fig:wilu_module}
\end{figure}

Fig. \ref{fig:wilu_module}a shows the execution of the WILU module. The WILU module reads the encoded and packed weight values from the weight BRAM as discussed in Section \ref{sec:meadow_arch}. A packet read from the weight BRAM contains mode bits and packed encoded weight values. The mode aware unpacking (MAU) module unpacks the packed encodings based on the mode as shown in Fig. \ref{fig:wilu_module}b. For example, for an 8-bit packed encoding, d0 to d7 is unpacked in 1, 2 and 4-bit values for modes 0, 1 and 2, respectively. The unpacked encodings are used to look up the reindexed unique matrix to get the actual weight values that are sent to the weight RF of the respective PE through the NoC.

\begin{figure*}[h!]
    \centering
    \includegraphics[width=\textwidth]{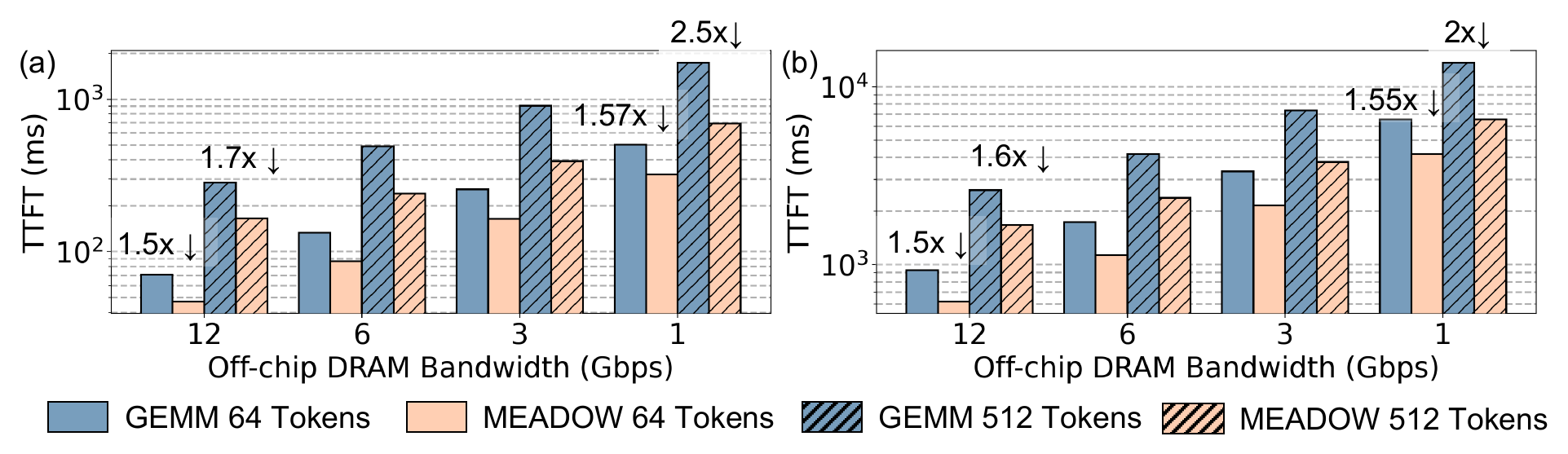}\vspace{-6mm}
    \caption{Time to first token (TTFT) Comparison of \ourwork with GEMM-based decoder implementation of the (a) OPT-125M and (b) OPT-1.3B LLM models on the ZCU102 FPGA with varying off-chip DRAM bandwidths. The evaluations are performed with 64 and 512 tokens during the prefill stage. }
    \label{fig:results_gemm_ours}
\end{figure*}

\begin{figure*}[h!]
    \centering
    \includegraphics[width=\linewidth]{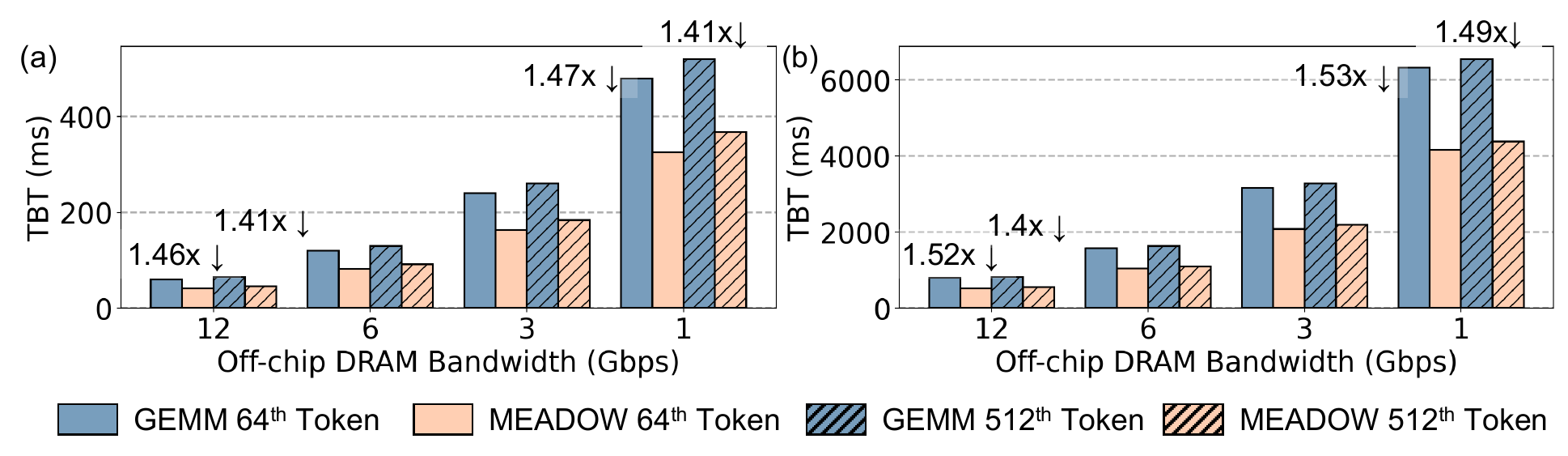} \vspace{-6mm}
    \caption{Time between tokens (TBT) comparison of \ourwork with GEMM-based decoder implementations of the (a) OPT-125M and (b) OPT-1.3B LLM models on the ZCU102 FPGA with varying off-chip DRAM bandwidths. For all cases the number of prefill tokens set to 512. The TBT is then measured for the 64th and 512th predicted token in the decode stage.}
    \label{fig:decode_gemm_ours}
\end{figure*}




\section{Results and Analyses}
\subsection{Experiment Setup}

\textbf{LLM Models and Datasets:} For benchmarking \ourwork, we use the OPT-125M and OPT-1.3B LLM models \cite{zhang2022opt} finetuned on the LAMBADA dataset using zero-shot adaptation with Smoothquant \cite{xiao2023smoothquant} post-training quantization. The weights and inputs are quantized to 8-bit precision. The 8-bit weight and input quantized OPT-125M and OPT-1.1B models achieve \textbf{60.7\% and 69.7\% accuracy} on the LAMBADA dataset.

\textbf{Xilinx ZCU102 FPGA Implementation:} For hardware evaluation, we implement the hybrid GEMM-Pipelined architecture of \ourwork on the Xilinx ZCU102 FPGA using the hardware parameters shown in Table \ref{tab:parameter_table}. The implementation uses 150K LUT, 845 BRAM and 2034 DSP resources. To maximize the number of PEs, we utilize both LUTs and the DSP blocks. Additionally, register files and pipeline registers are implemented using the LUT-based registers. 

\begin{table}[]
    \centering
    \resizebox{0.8\linewidth}{!}{\begin{tabular}{|c|c|}\hline
        \textbf{Parameter} & \textbf{Value} \\ \hline
         \#Parallel \& \#Broadcasting PEs & 84, 12 \\ \hline
         \#Multipliers per PE & 64 \\ \hline
         \#SM, \#LN \& \#ReLU Modules & 84, 8, 8 \\ \hline
         Weight, Input \& Output BRAM Size & 1MB, 1MB \& 1MB \\ \hline
         Weight, Input \& Output RF Size & 4KB, 4KB \& 4KB \\ \hline
         Clock Frequency & 100 MHz \\ \hline
    \end{tabular}}
    \caption{Hardware Parameter Table for ZCU102 FPGA Evaluation}\vspace{-4mm}
    \label{tab:parameter_table}
\end{table}

\textbf{GEMM Baseline:} To benchmark prefill and decode latency, we use the GEMM baseline. The GEMM baseline is realized by operating the \ourwork architecture in fully GEMM mode. Here, all the layers in the decoder \ttt{Q}, \ttt{K}, \ttt{V}, \ttsup{QK}{T}, \ttt{SMxV}, \ttt{Proj} and \ttt{MLP} are executed in the GEMM mode. This captures the standard execution pattern that is followed in all prior LLM optimization works. 

\textbf{Prefill and Decode Latency Measurement:} We use time to first token (TTFT) and time between tokens (TBT) to measure the prefill and decode latency, respectively. TTFT measures the time from when a prompt is submitted to the LLM until the first generated token is produced. It reflects the initial processing delay to infer the context of a given prompt by the LLM. TBT measures the latency of generating the $N^{th}$ token after the LLM has produced $N-1$ tokens post the prefill stage \cite{zhang2024llmcompass}. 

\textbf{\ourwork Operation Modes:} During the prefill and decode stage, we execute the TPHS dataflow for the \smqktv layers and GEMM is used for the remaining \ttt{K}, \ttt{V}, \ttt{Proj} and \ttt{MLP} layers. Weight Packing is applied in both stages. Note, during Decode, there is a marginal latency speedup with TPHS compared to GEMM operation for \smqktv since the input token size is 1. As we will see later, the decode stage latency gains are primarily stemming from weight packing.



\subsection{Prefill and Decode Latency Improvements}

\textbf{Prefill}: Fig. \ref{fig:results_gemm_ours}a and Fig. \ref{fig:results_gemm_ours}b compares the TTFT achieved by \ourwork and GEMM-based OPT-125M and OPT-1.3B LLM models for varying DRAM bandwidths. At DRAM bandwidth of 12 Gbps, \ourwork achieves 1.5$\times$-1.7$\times$ and 1.5-1.6$\times$ for OPT-125M and OPT-1.3B LLMs, respectively across different number of prefill tokens. At a low DRAM bandwidth of 1 Gbps, \ourwork achieves 1.57-2.5$\times$ and 1.55-2$\times$ lower TTFT compared to GEMM implementations for OPT-125M and OPT-1.3B LLM models, respectively. 

\textbf{Decode}: Fig. \ref{fig:decode_gemm_ours}a and Fig. \ref{fig:decode_gemm_ours}b compare the TBT achieved by \ourwork and GEMM-based approaches on the OPT-125M and OPT-1.3B LLM models, across varying DRAM bandwidths. For predicting the 64th and 512th token at 12 Gbps DRAM bandwidth, \ourwork reduces TBT by 1.4-1.46$\times$ and 1.4-1.52$\times$ for the OPT-125M and OPT-1.3B models, respectively. When operating at a constrained DRAM bandwidth of 1 Gbps, \ourwork achieves a 1.4$\times$-1.47$\times$ reduction in TBT for the OPT-125M model and a 1.5$\times$-1.53$\times$ reduction for the OPT-1.3B model.



\begin{figure}[h!]
    \centering
    \includegraphics[width=0.9\linewidth]{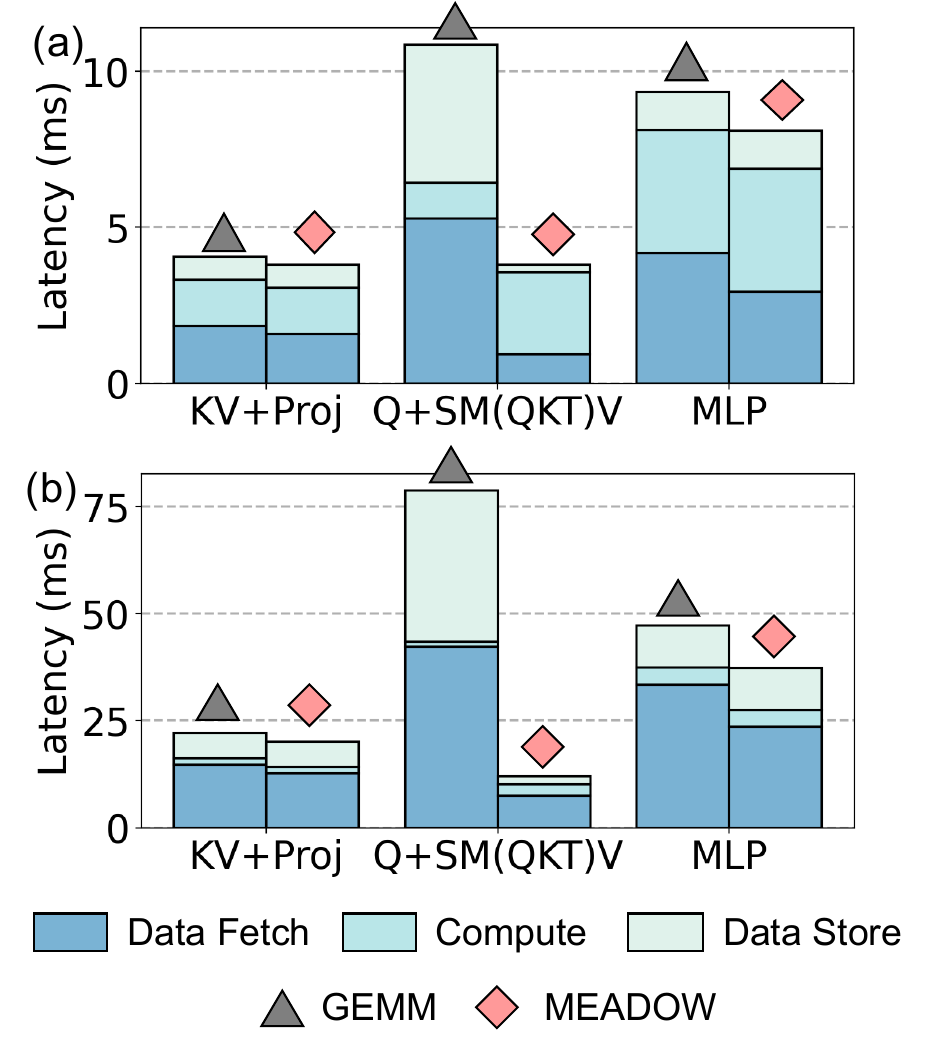}
    \caption{Prefill latency distribution for data fetch, compute and storage with 512 tokens at (a) 12 Gbps and (b) 1 Gbps off-chip DRAM bandwidth. The latency distribution is shown for one decoder layer of the OPT-125M LLM.}\vspace{-5mm}
    \label{fig:prefill_decode_distributions}
\end{figure}

The latency reduction observed in \ourwork for both prefill and decode stages stems from the targeted optimizations in data fetch and storage cycles. In GEMM-based implementations, executing the \ttt{Q+SM(QK\tsup{T})xV} layers during the prefill stage requires fetching weights and intermediate values from off-chip DRAM, performing matrix multiplications, and storing outputs back to DRAM. These data transfers impose substantial latency, especially as the size of the intermediate outputs scales directly with the number of attention heads and prefill stage tokens. This latency is exacerbated when the DRAM bandwidth is constrained (as illustrated in Fig. \ref{fig:prefill_decode_distributions}a and Fig. \ref{fig:prefill_decode_distributions}b).
 \ourwork's TPHS dataflow with pipelined operations within the \ttt{Q+SM(QK\tsup{T})xV} layers minimizes the number of off-chip memory accesses yielding a significant reduction in latency. For the \ttt{KV+Proj} and \ttt{MLP} layers, where data fetches are dominated by weight matrix transfers, the introduction of weight packing further reduces the latency by decreasing the volume of weight data fetched from the off-chip DRAM.


\begin{figure}
    \centering
    \includegraphics[width=0.9\linewidth]{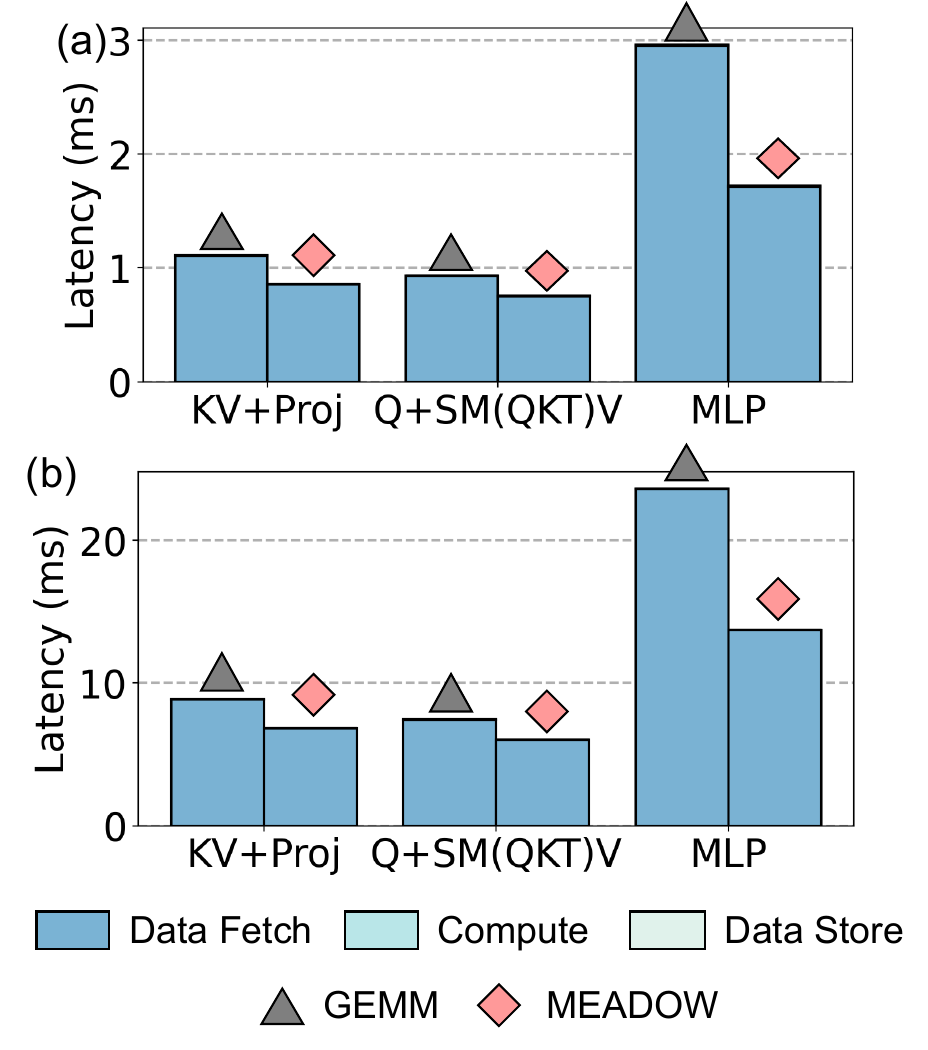}\vspace{-3mm}
    \caption{Decode latency distribution for data fetch, compute and storage at (a) 12 Gbps and (b) 1 Gbps off-chip DRAM bandwidth for one decoder layer of the OPT-125M LLM. The latency is shown for predicting the 64th token with 512 tokens at the prefill stage. The compute and store latencies are negligibly small compared to data fetch latency.}
    \label{fig:decode_distributions}
\end{figure}
During the decode stage, only a single token is processed at a time, significantly reducing input fetch and output storage demands compared to the prefill stage with its large pool of tokens. This limited data transfer, shown in Fig. \ref{fig:decode_distributions}a and Fig. \ref{fig:decode_distributions}b, makes weight data fetching the primary bottleneck. \ourwork is able to reach lower decode latency due to the weight packing strategy that reduces weight fetch latency.



\subsection{Efficacy of the Weight Packing Strategy}
\begin{figure}[h!]
    \centering
    \includegraphics[width=0.9\linewidth]{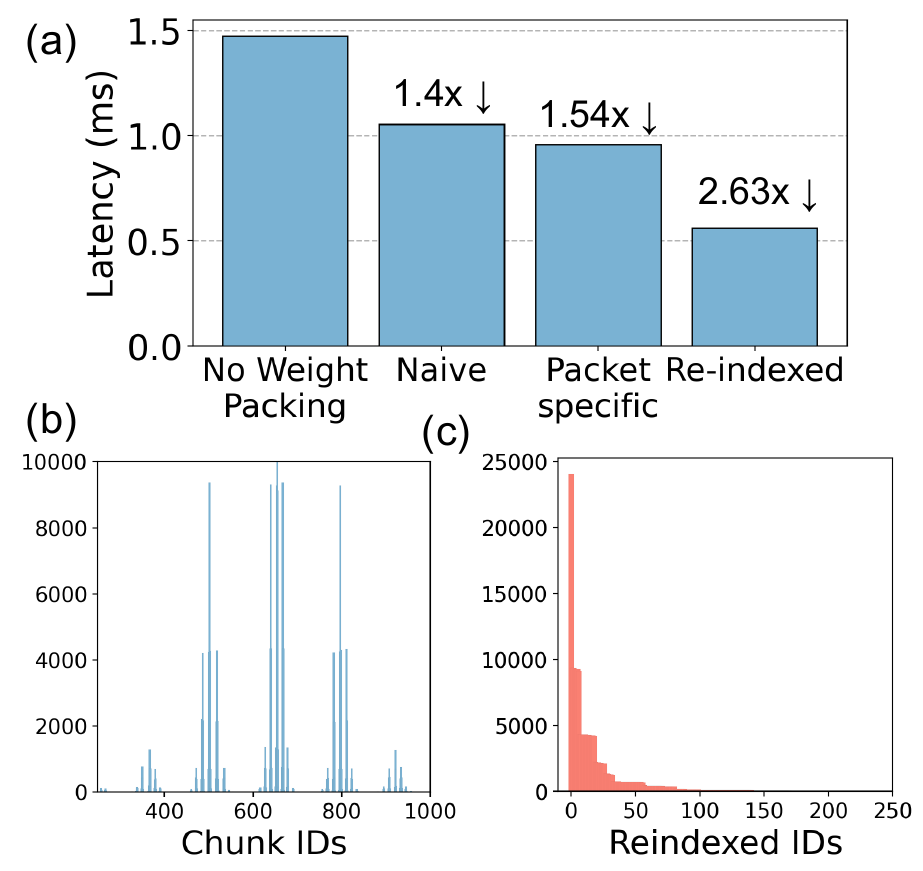}
    \caption{(a) Latency comparison of weight matrix transfer for 3 different weight packing optimizations. 1) indexing + naive data packing (Naive), 2) Indexing + packet specific encoding precision (Packet specific) and 3) frequency aware re-indexing + packet-specific encoding precision. (b) Histogram of the unique chunk IDs (shown for Chunk IDs between 200 and 1000) (c) histogram of the chunk IDs after performing frequency-aware re-indexing.}\vspace{-5mm}
    \label{fig:indexing_ablation}
\end{figure}
Indexing reduces a large weight matrix to unique chunk values and represents the weight matrix in terms of the unique chunk IDs. Fig. \ref{fig:indexing_ablation}a, analyses the latency improvements over different packing optimizations for the first MLP layer weights of decoder 1 of the OPT-125M LLM. The MLP1 weight is decomposed into 1272 unique chunks leading to a 11-bit encoded W precision. These 11-bit encoded W values are now grouped together into packets to improve the DRAM bandwidth efficiency.


As seen in Fig. \ref{fig:indexing_ablation}a, with naive packing, a latency improvement of merely 1.4$\times$ is achieved as several low-bit precision encoded W values are represented with 11-bit values. Upon using packet specific encoding precision, a 1.54$\times$ lower latency is observed as multiple low bit encoded W values are grouped per packet which reduces the number of data fetch cycles. 

The limited improvements in memory fetch latency with naive and packet-specific grouping arises due to the frequent occurrence of high-value chunk IDs, which hinders effective grouping of the encoded W values, as illustrated in Fig. \ref{fig:indexing_ablation}b. To this end, frequency-aware reindexing increases the number of low bit chunk IDs (Fig. \ref{fig:indexing_ablation}c) and thereby improves the packing efficiency leading to 2.63$\times$ lower weight fetch latency. 


\subsection{Comparison with Prior Works}
\begin{table}[h!]
    \centering
    \resizebox{\linewidth}{!}{\begin{tabular}{|c|c|c|c|}\hline 
         & CTA & FlightLLM  & \textbf{\ourwork}\\ 
         & \cite{wang2023cta} & \cite{zeng2024flightllm} & \textbf{(Ours)} \\ \hline
       KV, Proj, MLP  & GEMM & GEMM & \textbf{GEMM} \\ \hline
       Q, SM(QKT)V & GEMM & GEMM & \textbf{TPHS} \\ \hline
       Quantization & W8A8 & W8A8 & \textbf{W8A8} \\ \hline
       Weight Packing & \ding{55} & \ding{55} & \ding{52} \\ \hline
    \end{tabular}}
    \caption{Evaluation settings for prior work comparison. }
    \label{tab:prior_work}
\end{table}

We implement prior state-of-the-art LLM optimization approaches- CTA~\cite{wang2023cta} and FlightLLM \cite{zeng2024flightllm} on the \ourwork architecture with implementation parameters shown in 
Table \ref{tab:parameter_table}. As seen in Table \ref{tab:prior_work}, CTA \cite{wang2023cta} and FlightLLM \cite{zeng2024flightllm} execute all layers in the decoder in the GEMM mode. For fairness, the activations and weights in all works are maintained at 8-bit precision. \ourwork is implemented with weight packing and the TPHS dataflow for the \smqktv layers for both prefill and decode stages. The \ttt{K}, \ttt{V}, \ttt{Proj} and \ttt{MLP} layers are executed in the GEMM mode. 
\begin{figure}[h!]
    \centering
    \includegraphics[width=\linewidth]{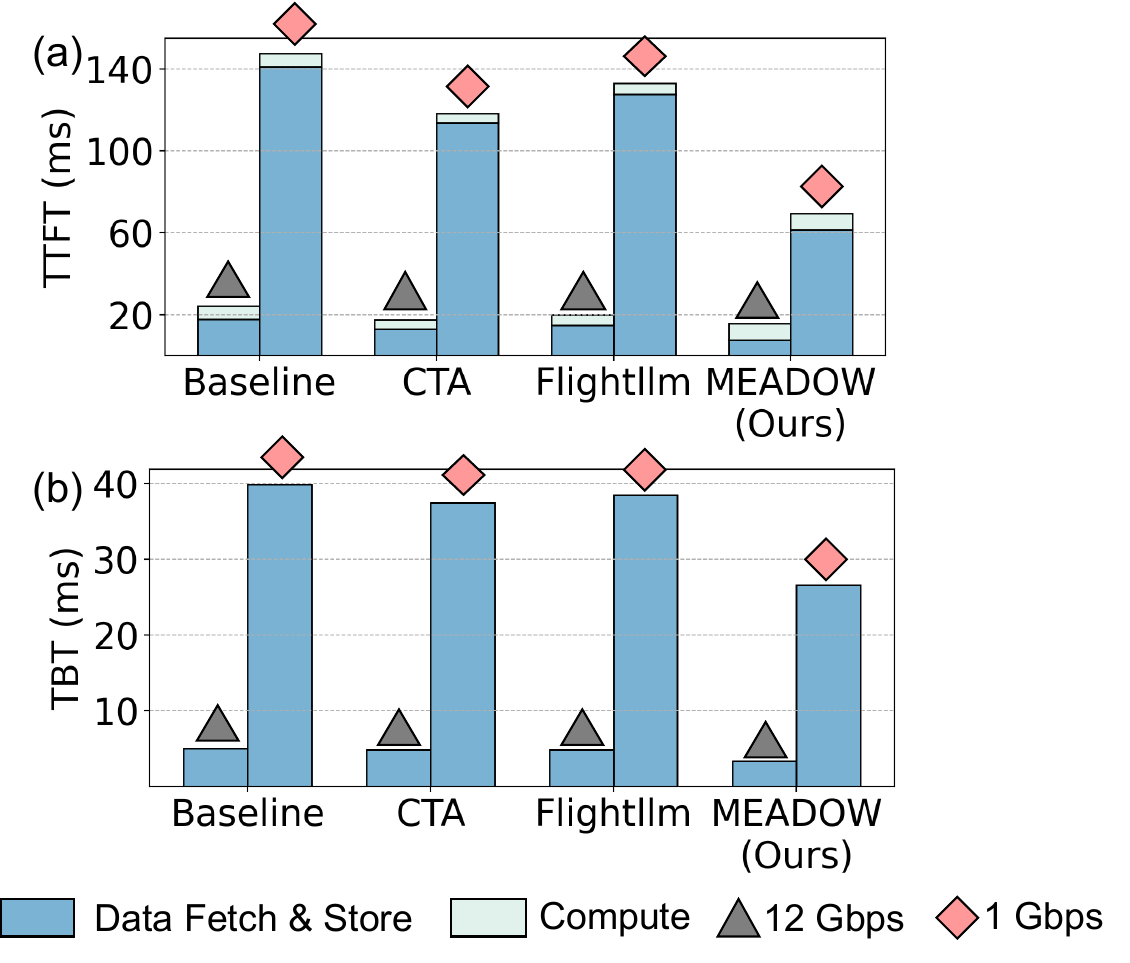}\vspace{-5mm}
    \caption{Figure comparing the (a) TTFT and (b) TBT latency of prior state-of-the-art LLM optimization works with \ourwork at different off-chip DRAM bandwidths.}
    \label{fig:prior_work_comp}
\end{figure}
Fig. \ref{fig:prior_work_comp}a and Fig. \ref{fig:prior_work_comp}b compares the TTFT and TBT latency of prior works with \ourwork. CTA \cite{wang2023cta} employs token compression to mitigate data redundancy, aiming to reduce memory and computational load by processing essential tokens only. While this approach decreases compute cycles, output storage, and input fetch latency in the \smqktv layers, the intermediate values for the remaining significant tokens still require fetching and storage in off-chip DRAM. Under constrained memory bandwidth, the latency involved in weight and token fetch/storage creates a substantial bottleneck, which limits CTA’s overall benefits during both prefill and decode stages.

FlightLLM \cite{zeng2024flightllm}, on the other hand, leverages unstructured N:M sparse acceleration architecture to cut down computations. While unstructured sparsity can lower compute requirements, it leaves input fetch latency largely unoptimized, and like CTA, FlightLLM does not apply any weight packing technique. To mitigate intermediate storage requirements during \smqktv operations, FlightLLM utilizes on-chip storage at decode time. However, since output storage latency during decode is negligible, as illustrated in Fig. \ref{fig:prior_work_comp}b, weights remain the dominant bottleneck, restricting overall performance gains.


Evidently, prior methods perform prefill and decode with unoptimized weight matrix sizes and only partially eliminate intermediate data fetch and storage cycles during the \smqktv operations. This partial approach limits their effectiveness, particularly under low-memory bandwidth constraints, where repeated fetches of intermediate values and weights cause latency bottlenecks. \ourwork offers architectural support and the TPHS dataflow innovation to completely eliminate the data fetch and storage latency of the \smqktv layers. Additionally, weight packing further reduces the latency of fetching the weight matrix. Overall, this translates to a 40\% improvement in the end-to-end latency with \ourwork compared to FlightLLM and CTA on ZCU102 FPGA-based OPT-125M implementation.


\subsection{Choosing between GEMM \& TPHS Dataflow}
\begin{figure}[h!]
    \centering
    \includegraphics[width=\linewidth]{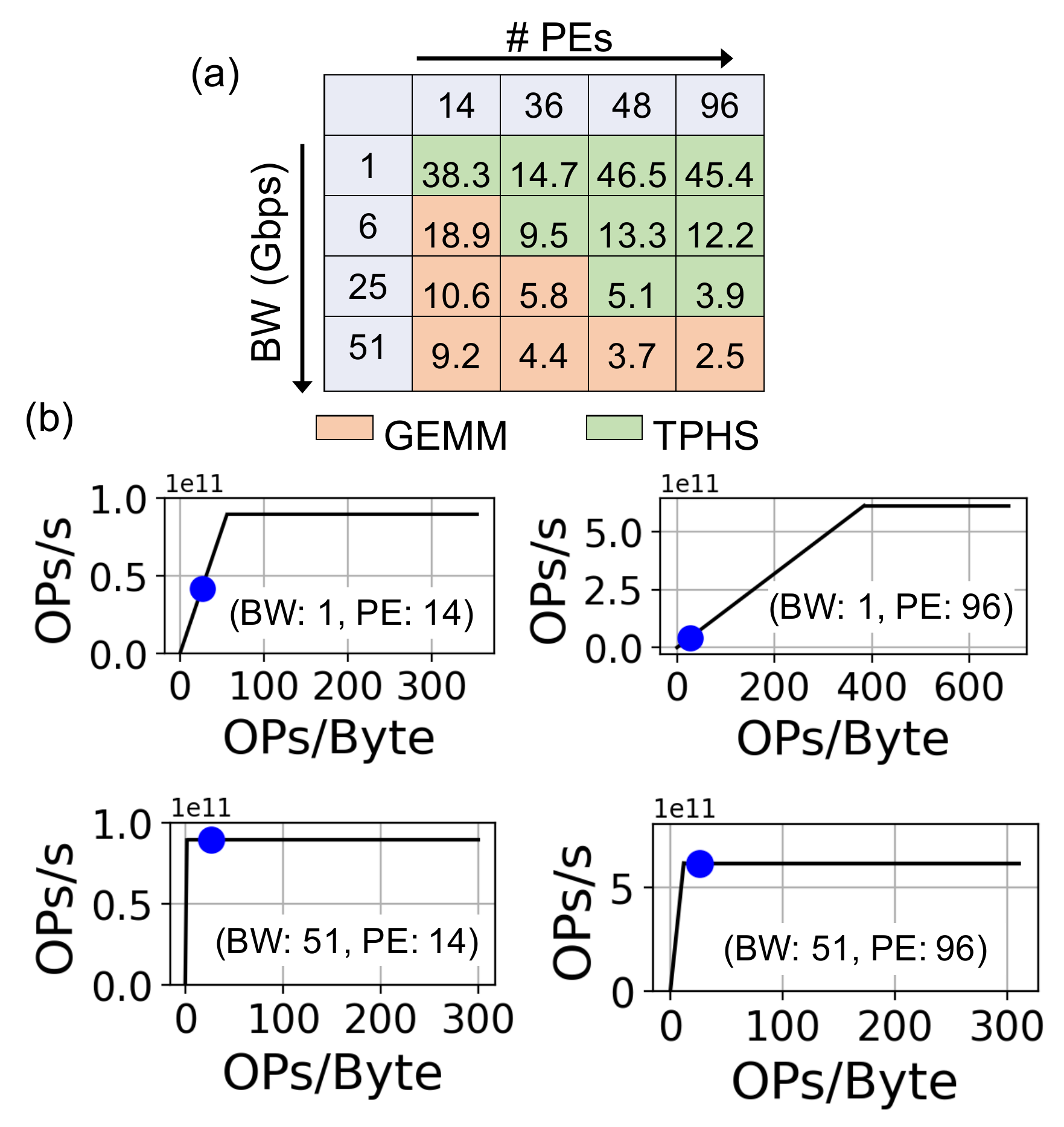}\vspace{-6mm}
    \caption{(a) Table showing optimal dataflow chosen for executing the \smqktv layers and the corresponding optimal prefill latency obtained for the OPT-125M LLM model. (b) The roofline plots for different (Bandwidth (BW), PE) configurations (1,14), (1,96), (51,14) and (51,96).}
    \label{fig:gemm_pipeline_suitability}
\end{figure}

From Fig. \ref{fig:gemm_pipeline_suitability}a, it is observed that the choice of GEMM and TPHS dataflow for the \smqktv layers is dependent on the number of PEs and the off-chip DRAM bandwidth. For high memory bandwidth scenarios, (BW:51, PE:14) and (BW:51, PE:96) GEMM is the dataflow choice. In contrast, TPHS is suitable for low memory bandwidth configurations (Fig. \ref{fig:gemm_pipeline_suitability}b). {This study justifies our framework as a suitable choice for deployment on a range of low memory capability edge devices.}

\subsection{ViT Latency Improvements with \ourwork}
\begin{figure}[h!]
    \centering
    \includegraphics[width=0.8\linewidth]{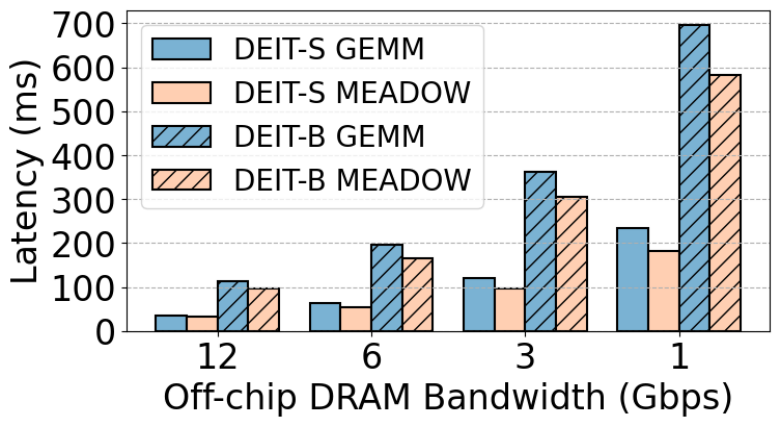}\vspace{-3mm}
    \caption{DeiT-S and DeiT-B ViT inference latency improvements with \ourwork compared to GEMM-based implementations on the ZCU102 FPGA. }\vspace{-2mm}
    \label{fig:enter-label}
\end{figure}

We also show the generality of \ourwork for ViT models. Vision transformers (ViTs) process multiple tokens together like the prefill stage of an LLM. With combined TPHS/GEMM dataflow and weight packing, \ourwork achieves 1.5-1.6$\times$ lower inference latency on the DeiT-S and DeiT-B \cite{touvron2021training} models trained on the ImageNet dataset \cite{deng2009imagenet} across different off-chip DRAM bandwidths. 

\section{Conclusion}
This work proposes \ourwork- targeting the latency intensive data fetch/store cycles of intermediate outputs and weights through the TPHS dataflow and Weight Packing to achieve 1.5$\times$ and 2.5$\times$ lower decode and prefill latency compared to GEMM-based implementations. \ourwork is crafted to achieve low latency LLM execution at highly constrained off-chip DRAM bandwidths achieving over 40\% end-to-end latency improvement compared to prior LLM optimization works. Additionally, we demonstrate the versatility of \ourwork by applying it towards ViT implementations. This typically makes \ourwork suitable for low power edge applications such as autonomous driving and mobile chatbots for both vision and NLP tasks.

\section{Acknowledgment}
This work was supported in part by CoCoSys, a JUMP2.0 center sponsored by DARPA and SRC, the National Science Foundation (CAREER Award, Grant \#2312366, Grant \#2318152), the DARPA Young Faculty Award and the DoE MMICC center SEA-CROGS (Award \#DE-SC0023198).



\bibliography{example_paper}
\bibliographystyle{mlsys2025}
\end{document}